\documentstyle[12pt]{article}

\font\sqi=cmssq8
\def\DR{\rm I\kern-1.45pt\rm R}
\def\DT{\rm {\bf T}}
\def\DC{\kern2pt {\hbox{\sqi I}}\kern-4.2pt\rm C}
\renewcommand{\thefootnote}{\fnsymbol{footnote}}
\begin{document}
\thispagestyle{empty}
{\hfill  IC/94/128}\vspace{2.5cm} \\
\begin{center}
{\rm International Atomic Energy Agency\\
and\\
United Nations Educational Scientific and Cultural Organization}\\
INTERNATIONAL CENTRE FOR THEORETICAL PHISICS\\
\vspace{2.5cm}
{\Large\bf ``Charge-Dyon" System As The Reduced
Oscillator } \vspace{1.5cm} \\

 {\large {\bf A.
Nersessian}\footnote{Supported in part by Grant No. M21000 from the
International Science
Foundation}\footnote{e-mail:nersess@theor.jinrc.dubna.su}}\vspace{0.5cm}\\
{\it International Centre for Theoretical Physics, Trieste, Italy}\\
\vspace{0.5cm}
and\\
\vspace{0.5cm}{\it Bogoliubov Theoretical Laboratory, JINR, Dubna,\\
Head Post Office,
P.O.Box 79, 101 000 Moscow, Russia}\footnote{Permanent address}\\
\vspace{1.5cm}
{\large{\bf V. Ter-Antonian}}\\
\vspace{0.5cm}{\it Bogoliubov Theoretical Laboratory, JINR, Dubna}\\
{\it Head Post Office, P.O. Box 79 , Moscow 101000, Russia} \end{center}
\vspace*{1cm}
\begin{abstract}
It is shown, that in the general case the Kustaanheimo-Stiefel
transformation (reduction of
$4d$ oscillator on the twistor space by the Hamiltonian action of $U(1)$
with further
time-reparametrization) leads  to the integrable system, describing
the interaction  of a
spinless electrically charged particle with a dyon .
\end{abstract}
\vfill
\begin{center}
MIRAMARE-TRIESTE\\
June 1994
\end{center}
\setcounter{page}0
\renewcommand{\thefootnote}{\arabic{footnote}}
\setcounter{footnote}0
\newpage
   \setcounter{equation}0
	    \section{Introduction }

The Coulomb system is one of the most important models of classical and
quantum mechanics. In spite of its simplicity, it has the
rich intrinsic structure, which is ensure by the non-kinematic
motion integral(s)
 --Runge-Lenz vector. The
existence of such an integral(s)
allows to obtain  Coulomb system from  the
reduction of
 $4d$ oscillator on twistor space
by the Hamiltonian action of $U(1)$ group and further
time-reparametrization
 (Kustaanheimo-Stiefel transformation, see, {\it e.g.} \cite{w}).
In the general
case Kustaanheimo-Stiefel transformation
leads to the more general integrable system \cite{i}.

In this paper we will re-derive this system, using twistor tecnique,
and show , that it describe  the {\bf interaction  of a spinless
electrically charged particle with  dyon} .\\
The paper is arranged as  follows.

In {\it Section 2}  the necessary information on the $4d$ oscillator
on  the twistor  space
 is given.

In {\it Section 3} this system is reduced by the Hamiltonian action of $U(1)$
group and  then lead its to the Coulomb-like form.
It is shown, that the obtained system describe the motion of spinless
electrically charged particle in the dyon field.\\

Below we will use  the relations
$\hat\sigma^{k}\hat\sigma^l =\delta^{kl}\hat E
                        +i\varepsilon^{klm}\hat\sigma^{m}$ and
 \begin{eqnarray}
(u\sigma^k\bar v )(r\sigma^k\bar s )& =& 2(u\bar s)(r\bar v)
-(u\bar v)(r\bar s) ,
\label{2.4}   \\
\quad i\varepsilon^{klm}(u\sigma^l\bar v
)(r\sigma^m\bar s )& =& (u\bar s) (r\sigma^k\bar v ) -(r\bar v )(u\sigma^k \bar
s )\label{2.3}
 \end{eqnarray}
where  $\hat\sigma^k$ ( {\scriptsize $ k, l, m$ }=$1, 2, 3$) are Pauli matrices
,
  $(u\bar v )\equiv u^A {\bar v}^A$ ,
 $(u\sigma^k\bar v )
\equiv u^A\sigma^{(k)}_{A\bar{B}}{\bar v}^B$,
$u^A$, $v^B$ are any pair of spinors ({\scriptsize$ A, B$} =$0, 1$ )
and  $"\bar{\quad}"$ denote complex conjugation  .
 \section{Oscillator}
 Let us consider the twistor space $\DT=\left(\DC^{4}, \Omega \right) $
with the  symplectic structure
 \begin{equation}
 \Omega =i (dz^{A}\wedge d{\bar\pi}^A +
d\pi^{A}\wedge d{\bar z}^A ),
 \label{1} \end{equation}
where  $ (\pi^A , z^A ) $ are the  complex coordinates (twistor) on $\DT$
(see {\it e. g.} \cite{p}, \cite{w}).\\
 The corresponding Poisson bracket    by the following nonzero basic
 relations defines :
 \begin{equation}
 \{z^A ,\bar\pi^B\}= \{\pi^A ,\bar z^B\}=
i\delta^{A\bar B}.\label{2} \end{equation}
 The Hamiltonian
 \begin{equation}
 H\equiv 2I^{0} =\omega\left((\pi\bar\pi)+ (z\bar z)\right)
\label{3} \end{equation}
 with the Poisson bracket  (\ref{2}) defines the dynamics of
$4d$ oscillator :
\begin{equation}
  \dot z^A =\{H , z^A\}=-i\omega\pi^A ,
\quad \dot \pi^A
=\{H , \pi^A\}=-i\omega z^A \label{4}.  \end{equation}
This system has the motion
integrals
\begin{eqnarray}
&I^{k}=\frac{\omega}{2}\left( (\pi\sigma^{k}\bar\pi ) +
(z\sigma^{k}\bar z ) \right),
\quad J^{k}=\frac{1}{2}\left( (z\sigma^{k}\bar\pi)
			+(\pi\sigma^{k}\bar
                        z)\right),&\label{5a}\\
&J^{0}=\frac{1}{2}\left( (z\bar\pi)
                        +(\pi\bar
                        z)\right),&\label{5c}
                         \end{eqnarray}
which define the  algebra :
\begin{eqnarray}
&\{J^k, J^l\}=\varepsilon^{klm}J^m ,\quad \{J^k, I^l\}=\varepsilon^{klm}I^m,
\quad \{I^k
, I^l\}=\omega^2\varepsilon^{klm}J^m,&\label{8a}\\
&\{J^k ,J^0\} = \{I^k ,J^0\} =0,\quad \{I^k ,I^0\}=\{J^k ,I^0\}=0.&\label{8b}
\end{eqnarray}
These  integrals  are not independent: taking into accunt
(\ref{2.4}), we obtain
$ \omega^2 (J^0)^2 = \omega^2 J^k J^k +I^k I^k
- ( I^0)^2. $

Suppose the Hamiltonian (\ref{3}) satisfies the initial condition
 $H=E_{\rm osc}$.
Thus the  equations of motion  (\ref{4})  can be  presented
in the form
  \begin{equation} \dot {\bf w}
=\{{\tilde H} , {\bf w}\} , \quad {\rm where }\quad {\tilde H}=H-E_{\rm osc}=0,
\quad{\bf w} =(\pi^A , z^A )
.  \label{13} \end{equation}
\section{Reduction}
It  follows from (\ref{8b}), that the motion integral $J^{0}$
commutes with
the functions   (\ref{5a}).
Therefore, the  algebra  (\ref{8a}) preserved  after   reduction of  the
 $4d$ system (\ref{4})  by the Hamiltonian action of $J^0$ .

 Let us fix the
(7-dimensional) level surfaces of $J^0$ :
 \begin{equation}
 J^0 =s.     \label{2.7}
 \end{equation}
The reduced phase space   ${\cal M}^{\rm red}$
 can be  obtained by the
   factorization of the level surface
(\ref{2.7})
    by the Hamiltonian action  of   $J^0$.

The set of 6 functionally independent functions on
$\DT$, which are invariant with respect to Hamiltonian action of $J_0$ ,
can be play the role of local coordinates on reduced phase space plays .
So the functions  $x^a =(q^k , p^k) $ (defined on the domain $r\equiv
(z\bar z)\neq 0$):
\begin{equation}
q^k = (z\sigma^k\bar z) ,\quad p^k=\frac{i}{2(z\bar
z)}\left((\pi\sigma^k\bar z )
-(z\sigma^k\bar\pi)\right) , \label{2.1a} \end{equation}
are the appropriate one $\{ x^a , J^0 \}= 0$ .

Since $J^0$ define  the $U(1)$ group  action on $\DT $ ,
the reduced
phase space is complex projective space: ${\cal M}^{\rm red} =\DC P(3)$,
{}.
 The inherited Poisson bracket on it
 by the following relation is define $$ \{f,g\}^{\rm
		       red}= \{f,g\}\bigg\vert_{J^0=s}, $$
       where $f,g$ are $U(1)$-invariant functions on $\DT$  .\\
 Using (\ref{2.4}), (\ref{2.3})
we get  :
\begin{equation}
\{q^k , q^l \}^{\rm red}=0 ,
			\quad  \{p^k , q^l \}^{\rm red}= \delta^{kl} ,\quad
   \{p^k , p^l \}^{\rm red} =
-s\varepsilon^{klm}\frac{q^m}{r^3} ,\label{2.2}
 \end{equation}
where $r =\sqrt{q^k q^k}$ .\\
This is the Poisson bracket on the twisted cotangent bundle .
It describes the motion of the
{\it  electrically charged particle in the
  magnetic monopole field.}

The  $U(1)$-invariant functions on $\DT$ by the restriction on the
  level surface (\ref{2.7}) can be reduced on the
${\cal M}^{\rm red}$ .
Thus
the   reduced  Hamiltonian  and  the motion  integrals
  (\ref{5a})  after re-scaling ($q^k\to 2m\omega q^k,
\;\;p^k\to \frac{p^k}{2m\omega}$) take the form :
 \begin{equation}
{\tilde H}_{\rm red}=(H-E_{\rm osc} )\bigg\vert_{J_0=B}=
 r\left(H_{\rm red}+2m\omega^2\right) =0,     \label{in}
\end{equation}
where
\begin{equation}
 H_{\rm red}= \frac{p^k p^k}{2m}- \frac{E_{\rm osc}}{r} +\frac{s^2}{2mr^2}
\label{c}   , \end{equation}
  \begin{eqnarray}
& J^k_{\rm red}&=J^k\bigg\vert_{J^0=s}=
\varepsilon^{klm}p^l q^m +\frac{sq^k}{r} -{\rm total}\;\;{\rm angular}
\;\;{\rm momentum},\label{2.11a}\\
& I^k_{\rm red}&= I^k\bigg\vert_{J^0=s}=\frac{1}{2m}\varepsilon^{klm}p^l
J^m_{\rm red}
 +\frac{q^k E_{\rm osc}}{2r} +\nonumber\\
&\quad&\quad\quad\quad\quad\quad + mq^k \left( H_{\rm
red}+2m\omega^2\right)-{\rm
Runge- Lenz}\;\;{\rm vector}.\label{2.11b} \end{eqnarray}
Therefore the Hamiltonian system (\ref{4}) is reduced to
\begin{equation}
\frac{dx^a}{rdt}=
\{ H_{\rm red}, x^a\}^{\rm red},
 \label{17} \end{equation}
with the initial condition (\ref{in}), Poisson bracket (\ref{2.2}) and the
motion integrals (\ref{2.11a}), (\ref{2.11b}). \\
The last term in (\ref{2.11b}) obviously commutes with $H_{\rm
red}$, $I^k_{\rm red}, J^k_{\rm red} $ and vanishes
because the initial condition (\ref{in}). So it can be omitted.
\\

This system permits simple physical interpretation:
it describes the
motion of the  {\it electrically charged
spinless particle  in the dyon field
}, or,
equivalently, their relative motion in "centre-of-mass" systems.\\
Indeed, let we have the dyon with magnetic charge $g$
and electric charge $Q$ in the origin of the coordinates
 and
the spinless particle with the electric charge $e$ in its  field.
The electrostatic interaction of these particles is defined by
the second term
 of Hamiltonian  (\ref{c}), and the interaction with the monopole's
 magnetic field, by the Poisson bracket (\ref{2.2}), so
$$eQ=E_{\rm osc},\quad \frac{eg}{c}=s. $$
The of the electron-monopole
system has its intrinsic angular momentum (spin) ({\it e.g.} \cite{gold})
 $${\vec s}
=\frac{eg}{c}\frac{{\vec
r}}{r}$$
and, indeed, it is presented  in (\ref{2.11a}) .
Correspondingly,  the last
term in the Hamiltonian (\ref{c}) defines  the interaction  of
the system's own magnetic momentum ${\vec\mu}$ with the
dyon magnetic field  ${\vec H}$ :
$$\frac{s^2}{2mr^2}=({\vec\mu}{\vec H}) , \quad{\rm where}\quad
{\vec\mu} =\frac{e{\vec s}}{2mc}, \;\;\;\;{\vec H} =\frac{g{\vec
r}}{r^3}.$$

So, the obtained integrable system correctly describes the
``charge -- dyon"  interaction.

It is interesting to consider
the correspondence between oscillator and this system on the
{\it quantum } level \cite{m}, namely,  canonically  quantize the
oscillator (in correspondence with  (\ref{2}))
and then reduced it in accordance with the above-presented scheme.
 For example,
the representation
$${\hat z}^A = z^A,\quad
 {\hat{\bar\pi}}^A =\hbar\frac{\partial}{\partial z^A},\quad {\hat{\bar
z}}^A ={\bar z}^A ,\quad {\hat\pi}^A =-
\hbar\frac{\partial}{\partial {\bar z}^A}$$
is appropriate for  such a reduction .

One can attempt to construct in the same manner the Hamiltonian
system describing the
motion of {\it spinning particle} in the dyon field,
 reducing the oscillator on the
 super-twistor space , using the approach, developed in
 \cite{knjmp2}.
\section{Acknowledgment}
One of us (A.N.) is very indebted to Prof. S. Randjbar-Daemi for the
helpful discussions and Prof. Abdus Salam, the International Atomic Energy
Agency and UNESCO, for  hospitality at International Centre for Theoretical
Physics, Trieste
where the work has been completed.

 \begin{thebibliography}{33}
 \bibitem{w}  N.M.J. Woodhouse --{\it Geometric Quantization.},
Clarendon Press, Oxford, 1992
\bibitem{p} R.Penrose, W. Rindler --{\it Spinors and space-
time. Vol. 2: Spinor and twistor

methods in space-time geometry.}, Cambridge University Press, Cambridge, 1986
\bibitem{i}T. Iwayi, Y. Uwano -
 J.Math.Phys., {\bf 27} (1986), 1523

\bibitem{gold}A.Goldhaber - Phys.Rev., {\bf B140} (1965), 1407
\bibitem{m} V.  Mladenov , V . Tsanov-- J. Physics {\bf A20} (1987),5865\\
I. M. Mladenov--Ann. Inst. H. Poincare, Phys. Theor., {\bf 50}(1989), 183

\bibitem{knjmp2} O.M.Khudaverdian, A.P.Nersessian -
 J.Math.Phys., {\bf 34} (1993), 5533